\documentclass[a4paper,fleqn,usenatbib,useAMS]{mnras}
\usepackage[final]{graphicx}
\usepackage{amssymb}
\usepackage{amsmath}
\usepackage{lastpage}
\usepackage{dsfont}
\usepackage{float}
\usepackage{caption}

\usepackage{subcaption}
\usepackage{rotating}

\usepackage{tabularx}
\usepackage{pdfpages}
\usepackage{color}
\usepackage[toc,page]{appendix}
\usepackage{gensymb}
\usepackage{epstopdf, epsfig}
\usepackage{hyperref}
\usepackage{pgfplots}
\pgfplotsset{compat=newest}
\usepackage{multicol}  
\usepackage{bm}	
\usepackage{pdflscape}
\usepackage[T1]{fontenc}
\usepackage{ae,aecompl}
\usepackage{pdflscape}

\title[Effects of capturing a wide-orbit planet on planetary systems]{Effects of capturing a wide-orbit planet on planetary systems: system stability and Habitable Zone bombardment rates}
\date{\today}
\author[G. Kokaia, M. B. Davies, A. J. Mustill]{Giorgi Kokaia$^{1}$\thanks{Contact e-mail: \href{mailto:giorgi@astro.lu.se}{giorgi@astro.lu.se}}, Melvyn B. Davies$^{1}$, Alexander J. Mustill$^{1}$
\\
$^{1}$Lund Observatory, Department of Astronomy and Theoretical Physics, Lund University, Box 43, SE-221 00 Lund, Sweden}

\date{\today}

\pubyear{2020}

\begin{document}
\label{firstpage}
\pagerange{\pageref{firstpage}--\pageref{lastpage}}
\maketitle

\begin{abstract}
A large fraction of stars are formed in dense clusters. In the cluster, close encounters between stars at distances less than 100 au are common. It has been shown that during close encounters planets can transfer between stars. Such captured planets will be on different orbits compared to planets formed in the system, often on very wide, eccentric and inclined orbits. We examine how these captured planets affect Kuiper-belt like asteroid belts in their new systems, and how this affects habitable planets in the system. We show that these captured planets can destabilize the asteroid belt, and we show that the fraction of the asteroid that make it past the giant planets into the system to impact the habitable planet is independent of the captured planets orbital plane, whereas the fraction of the asteroids that are removed and the rate at which they are removed depend strongly on the captured planets pericentre and inclination. We then examine all possible outcomes of planet capture and find that when a Jupiter-mass planet is captured it will in 40\% of cases destabilize the planets in the system, in 40\% of cases deplete the asteroid belt in a few Myr, i.e. not posing much risk to life on terrestrial planets which would be expected to develop later. In the final 20\% of cases the result will be a flux of impactors 5-10 times greater than that on Earth that can persist for several Gyr, quite detrimental to the development of life on the planet.
  
\end{abstract}
\section{Introduction}
The majority of stars form in clusters~\citep{Lada2003}. These clusters have stellar densities of 100--1000 per ${\rm pc}^{-3}$ ~\citep{Dias2002, Lamers2005}, compared to $\sim1$ per ${\rm pc}^{-3}$ in the field (see e.g.~\citealp{GaiaDR2}), and lifetimes of several tens to hundreds of Myr~\citep{Gieles2007}. In such dense clusters 10--20\% of Solar-like stars will experience an encounter within 100 au with another star, which is close enough to affect the stability of existing planets~\citep{Malmberg2011}. 

Given that the giant planets form before gas dispersal in proto-planetary discs, typically within a few Myr after star formation~\citep{Machida2010,Piso2014,Piso2015,Bitsch2019}, for most of these encounters the stars will have finished forming their giant planets.
During these encounters, planets can be transfered between stars~\citep{Li2019}. This can result in planets on orbits that are inconsistent with planet formation models. This is a possible explanation for the hypothesised presence of Planet 9~\citep{Batygin2016} on its (albeit loosely constrained~\citealp{Brown2016}) few hundred au wide orbit~\citep{Mustill2016}.

In this paper we will examine how capturing a planet will affect planets in the habitable zone. In particular we consider systems like our own, i.e. a terrestrial planet at $\sim$1 au, with giant planets beyond it surrounded by a Kuiper-belt like asteroid belt (throughout the paper we refer to this belt simply as the ``asteroid belt" and its component bodies as ``asteroids"). There are four ways in which the captured planet can affect them:
\begin{enumerate}
    \item It can be captured on an orbit that crosses the orbits of planets in the system, meaning it can either directly scatter a planet in the habitable zone or trigger an instability amongst the outer planets. Instabilities amongst the outer planets can destabilize a large fraction of possible orbits in the habitable zone~\citep{Carrera2016, Kokaia2019b}.

    \item The captured planet can induce Kozai oscillations, which can destabilize all of the planets without the captured planet ever directly interacting with them. The stability of planetary systems with an external highly inclined companion has been well studied for both stellar and planetary companions by e.g~\cite{Innanen1997, Kaib2011, Malmberg2011, Mustill2017, Denham2019}.

    \item  In the absence of an instability the outer planets or a single outer planet can have its eccentricity pumped up putting it on an orbit crossing the habitable zone which will destabilize most if not all possible orbits within it~\citep{Jones2001, Menou2003, Matsumura2013, Agnew2018, Georgakarakos2018}.

    \item Asteroids in belts external to the planets such as the ones found in the Kuiper belt (which appears to be fairly common around Sun-like stars~\citealp{Montesinos2016, Sibthorpe2018}) can be directly scattered or undergo Kozai-oscillations and end up crossing the orbits of the giant planets in the system. They can then be scattered further inwards and end up on orbits crossing the habitable zone. This can then lead to a large number of asteroid-impacts on any planets in the habitable zone.

\end{enumerate}

The main focus of the paper will be outcome iv), with outcome i) discussed briefly. In section~\ref{sec:kz} we will discuss the theoretical framework of the Kozai mechanism and also show the results of an exploratory experiment comparing the theory with our simulation setup. This setup will be discussed in section~\ref{sec:setup} and the findings from it will be presented in section~\ref{sec:result}. In section~\ref{sec:disc} we discuss the implications of said findings and also explore the impact of having different setups.

\section{The Kozai mechanism}\label{sec:kz}

The Kozai mechanism~\citep{Kozai1962, Lidov1962} is a secular three-body effect which can produce very large oscillations in eccentricity and inclination. Here follows a brief description of the process: Consider an inner body (A) and an outer companion (C) orbiting a central body ($\star$). A particularly simple case of Kozai cycles occurs when the three conditions below are met~\citep{Innanen1997, Malmberg2007b}.
\begin{itemize}
    \item $a_C\gg a_A$: The semi-major axis of C needs to be much larger than the semi-major axis of A.
    \item $m_C\gg m_A$: The mass of C needs to be much larger than the mass of A.
    \item $180^\circ-i_{\rm crit}>i>i_{\rm crit}$: The mutual inclination ($i$) between A and C need be between the the critical angle $i_{\rm crit}$ and $180^\circ-i_{\rm crit}$ where $i_{\rm crit}=\arcsin\sqrt{2/5}\approx39.23^\circ$.
\end{itemize}
Cycles in eccentricity and inclination will still happen if either of the first two conditions are not met. However, the formalism that follows will be less and less accurate as the mass and semi-major axis ratios approach unity~\citep{Katz2011, Naoz2013a, Naoz2013b, Antognini2015}.

\begin{figure}
\centering
\includegraphics[width = 1\columnwidth]{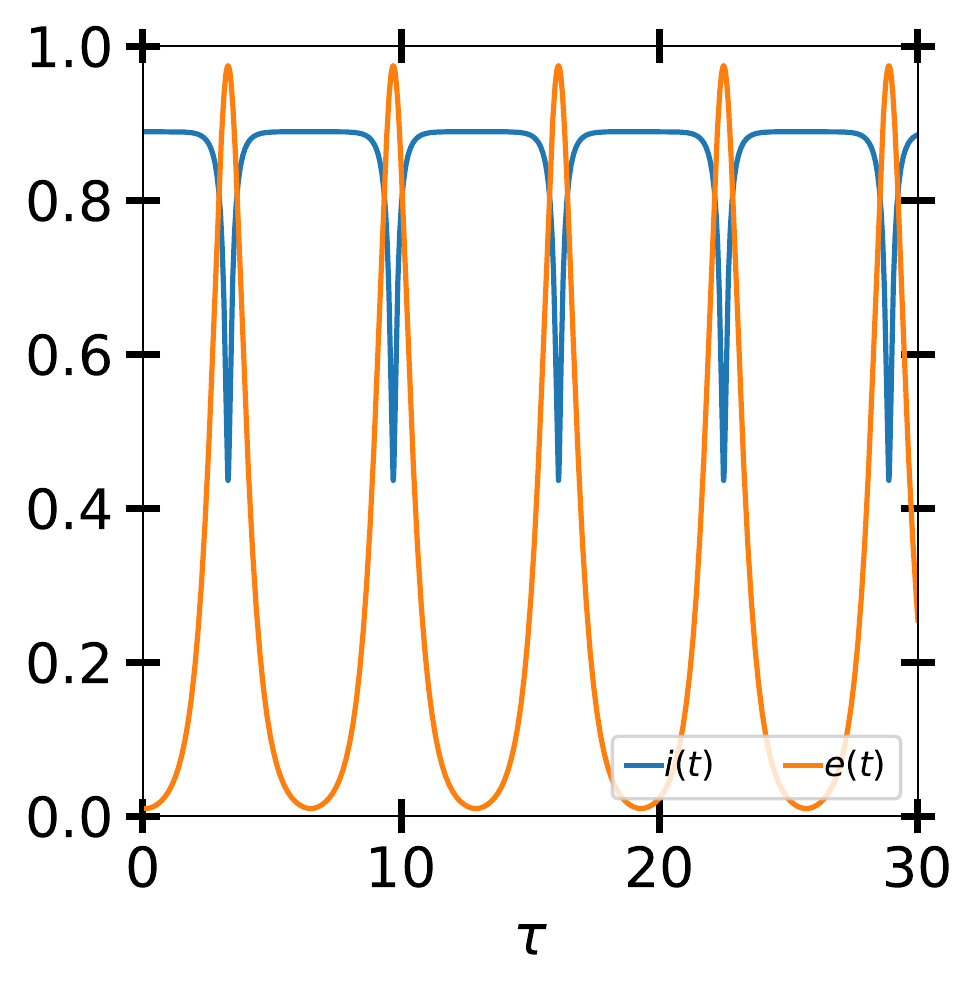}
\caption{The eccentricity and inclination evolution as given by equation~\ref{eq:Kozai} with initial values of $e_0=0.01$ and $i_0=80^\circ$. The inclination has been re-normalized such that $90^\circ=1$, the maximum eccentricity reached is $e=0.97$ and is given by equation~\ref{eq:e_max}.}
\label{fig:kozai}
\end{figure}

The set of differential equations (as given in~\citealp{Valtonen2006}, which is taken from~\citealp{Innanen1997} but correcting a sign error in the differential equation describing the inclination) describing the evolution of the Keplerian orbital elements [$i$: inclination, $e$: eccentricity, $\omega$: argument of pericentre, $\Omega$: longitude of ascending node] of the body undergoing the oscillations is shown below:
\begin{equation}
    \begin{array}{lll}
        \dfrac{di}{d\tau} & = & -\dfrac{15}{8}\dfrac{e^2}{\sqrt{1-e^2}}\sin2\omega\sin i\cos i,  \\\\
        \dfrac{de}{d\tau} & = & \dfrac{15}{8}e\sqrt{1-e^2}\sin2\omega\sin^2i, \\\\
        \dfrac{d\omega}{d\tau} & = & \dfrac{3}{4}\dfrac{1}{\sqrt{1-e^2}}\left[2(1-e^2)+5\sin^2\omega(e^2-\sin^2i) \right], \\\\
        \dfrac{d\Omega}{d\tau} & = & -\dfrac{\cos i}{4\sqrt{1-e^2}}(3+12e^2-15e^2\cos^2\omega).
    \end{array}
    \label{eq:Kozai}
\end{equation}
with $l$,
\begin{equation}
    l = \sqrt{1-e^2}\cos i,
    \label{eq:conserved}
\end{equation}
as a conserved quantity, which is a measure of how much angular momentum can be redistributed in the system. Here $i$ is the mutual inclination between A and C, $\tau$ is a re-scaling of the time, which for C with semi-minor axis $b_\mathrm{C}$ is given as:
\begin{equation}
    \tau = \dfrac{a_\mathrm{A}^{1.5}}{b_\mathrm{C}^3}\dfrac{m_\mathrm{A}}{\sqrt{GM_\star}}t,
    \label{eq:tau_t}
\end{equation}
where $G$ is the gravitational constant and $M_\star$ the stellar mass. From this we can determine the Kozai time-scale as a function of the period of A ($P_{\rm A}$) as:
\begin{equation}
    T_\mathrm{Kozai}\sim P_\mathrm{A}\left(\dfrac{a_\mathrm{C}}{a_\mathrm{A}}\right)^3\left(\dfrac{M_\star}{m_\mathrm{C}}\right)(1-e_\mathrm{C}^2)^{3/2}
    \label{eq:Ktimescale}
\end{equation}
We integrate the equations and show the evolution of eccentricity and inclination in figure~\ref{fig:kozai}, where we see the oscillations as a function of the time parameter $\tau$. Given the conserved quantity, $l$, we see that as expected A reaches its maximum eccentricity when it is at its minimum inclination. We can determine the maximum eccentricity by setting $\cos i_A=\cos i_{\rm crit}$ in equation~\ref{eq:conserved}.
\begin{equation}
    e_\mathrm{max} = \sqrt{1-(5/3)(1-e_0^2)\cos^2 i_0}
    \label{eq:e_max}
\end{equation}
where $e_0$ and $i_0$ are the initial values for the eccentricity of the inner body and mutual inclination between inner and outer body.
\subsection{Kozai between planets}{\label{sec:pp_Kz}}
The Kozai mechanism described in the previous section details how the system behaves in the test particle limit, i.e. when all the angular momentum is held by the inclined perturber. 

The system which has been explored in greatest detail throughout this paper is set up as follows: a central body of one Solar mass, an Earth--mass planet at 1 au, a Jupiter--mass planet on a circular orbit at 10 au and an additional Jupiter--mass planet on a circular but inclined orbit at 100 au. We refer to each body as Sun, Earth, [lower case] jupiter and Companion.

\begin{figure}
    \centering
    \includegraphics[width = \columnwidth]{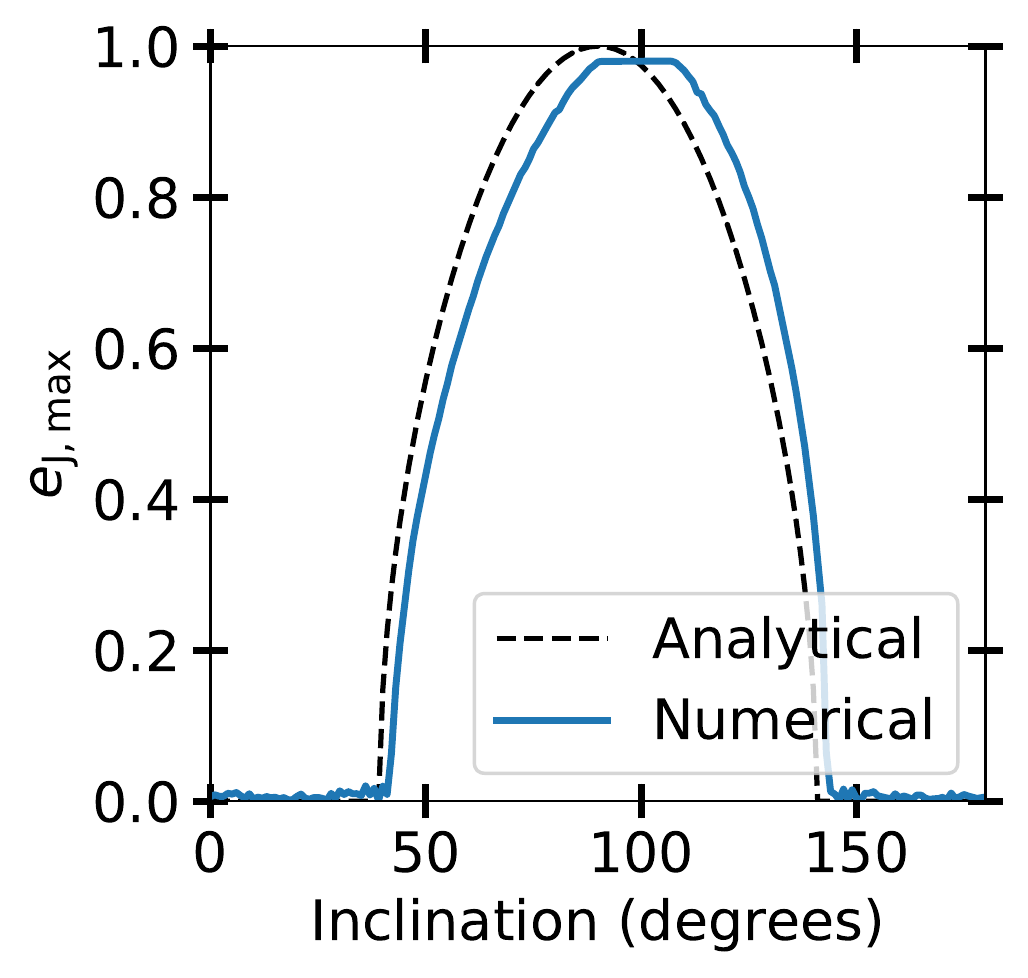}
    \caption{A range of simulations for the SEjC system. The dashed black line shows the theoretical $e_{\rm max}$ predicted from equation~\ref{eq:e_max}, i.e. the test particle limit. The solid blue line shows the maximum eccentricity achieved by the jupiter in the SEjC runs.}
    \label{fig:kztest}
\end{figure}

We investigate how the behaviour of this system differs from what we expect from the analytical solution above by running a set of simulations for just these four bodies (SEjC) where we vary the inclination of the Companion ($i_C$) from 0 to 180\footnote{Inclinations for the Companion above 90 degrees denote retrograde orbits.} degrees, with both the Earth and jupiter in the reference plane. In figure~\ref{fig:kztest} we show the maximum eccentricity reached by jupiter as a function of eccentricity. The black dashed line shows the theoretical expectation, which is symmetric about 90 degrees as expected; given equation~\ref{eq:e_max}. The solid blue line which shows the simulated data is no longer symmetric due to the angular momentum of jupiter and the location of the maximum has now shifted from 90 to $\sim$100 degrees. The flat peak of the blue line is due to the Sun having a radius of 0.2 au in the simulations, i.e. when anything gets within said distance it is removed. This was chosen for computational efficiency in the runs with asteroids later one, but we chose to keep it as such here for consistency.

This means that jupiter reaches an eccentricity of 0.98 for $i_C$ between 90 and 110 degrees, rather than between 81 and 99 degrees indicating an asymmetry between pro and retro-grade orbits. This makes sense because jupiter holds $\sim$1/3 of the system's angular momentum and thus putting the Companion on a retrograde orbit will not invert the total angular momentum-vector of the system.  

\subsection{Time-scales}
As we will see later, time-scales will prove to be an important aspect of what happens when a planet is captured. This is because a planet on an orbit that is precessing sufficiently fast will not have its eccentricity pumped up by the Kozai mechanism~\citep{Kaib2011, Mustill2017, Denham2019}. Orbits precess in any non-Keplerian potential (see e.g.~\citealp{Murray1999}), i.e. any two bodies orbiting a central more massive body will precess.

This means that if there are multiple planets in the system their orbits can be such that they will  not undergo Kozai cycles. For precession to prevent Kozai cycles it needs to happen on shorter time-scales than the Kozai time-scale. The precession time-scale on an orbit exterior to the orbit of $N_\mathrm{inner}$ number of planets is given by~\citep{Mustill2017}:
\begin{equation}
    T_\mathrm{prec}\sim\frac{4}{3} P_\mathrm{outer}\left(\sum_{i=1}^{N_\mathrm{inner}}\frac{m_ia_i^2}{M_\star a_\mathrm{outer}^2}\right)^{-1}.
    \label{eq:Ptimescale}
\end{equation}

In the setup we are looking at we then find four time-scales: $T_\mathrm{P,p}$ -- the precession time-scale in the planet--planet interaction; $T_\mathrm{K,a}$, $T_\mathrm{K,p}$ -- the Kozai time-scales of the asteroid belt and planets given by equation~\ref{eq:Ktimescale}; and $T_\mathrm{P,a}$ -- the precession time-scale of the planets acting on the asteroid belt. In order to avoid system-wide instability, while allowing for delivery of asteroids to the HZ, these need to satisfy $T_\mathrm{P,p}< T_\mathrm{K,p}< T_\mathrm{K,a}< T_\mathrm{P,a}$. Given the scaling with semi-major axis in eqs.~\ref{eq:Ktimescale} and~\ref{eq:Ptimescale} we can ignore $T_\mathrm{K,p}$ as it will always be smaller than $T_\mathrm{K,a}$ if that is smaller than $T_\mathrm{P,a}$. That leaves us with the time-scale condition:

\begin{equation}
    T_\mathrm{P,p}< T_\mathrm{K,a}< T_\mathrm{P,a}
    \label{eq:condition}
\end{equation}

However it should be noted that, for most planetary systems containing more than one planet the first inequality will be satisfied if the second one is.

\section{Setup of the main experiments}\label{sec:setup}

All the simulations were performed using the \textsc{mercury} N-body package~\citep{Chambers1999}. We used the Bulirsch-Stoer integrator with an accuracy parameter of $10^{-12}$ resulting in energy errors of $<10^{-6}$. We set an ejection radius of 1000 au and  set a stellar radius of 0.2 au. Both are done to speed up the simulations (reducing the number of bodies simulated and avoiding small pericentre passages). The stellar radius is also motivated physically as the asteroids (test particles) simulated are icy bodies and would likely not survive such a pericentre passage.

In all the simulations containing an Earth we inflated it to 100 Earth radii in order to have some asteroids hitting the body without having to use an unfeasible number of test particles. Throughout the paper we will use a dagger ($\dagger$) to differentiate hitting the inflated Earth from hitting an actual Earth-sized Earth.

\subsection{Asteroid belt}

To the SEjC system we now add asteroids and examine how they interact with the Companion and with jupiter. For this we use an asteroid-belt analogous to the Kuiper-belt using the prescription from~\cite{Volk2011}.

\begin{figure}
\centering
\includegraphics[width = 1\columnwidth]{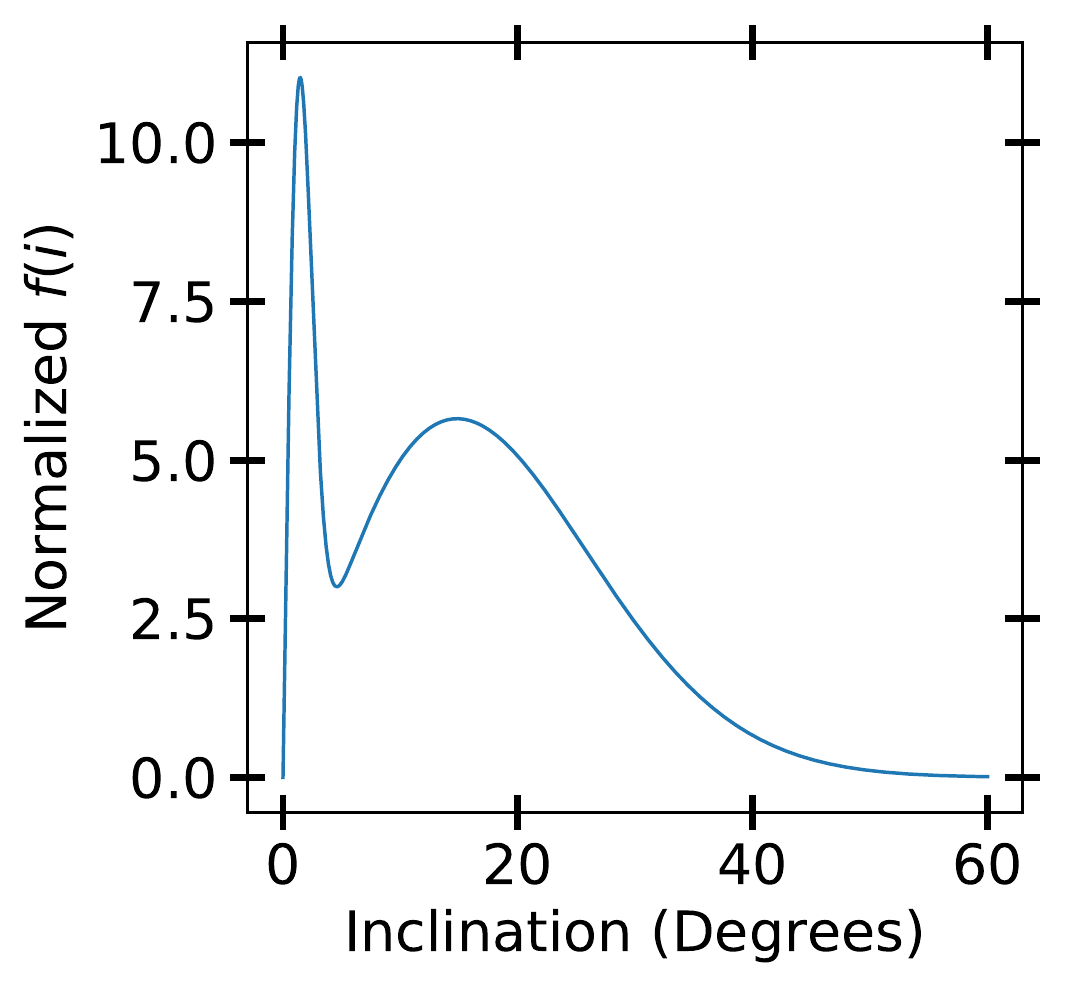}
\caption{The two-component inclination distribution of the Kuiper-belt, as determined in~\protect\cite{Volk2011}.}
\label{fig:inc_init}
\end{figure}

In the prescription the asteroids have a uniform semi-major axis distribution between 40 and 50 au. The eccentricity distribution is a Gaussian with $\mu=0.15$ and $\sigma=0.07$ truncated at 0 and 0.3, putting their orbits between 28 and 65 au. The inclination of the asteroids is modelled with a hot and a cold component, the distribution is given below:

\begin{equation}
    f(i)=\sin i\left[A\exp\left(\frac{-i^2}{2\sigma_1^2}\right)+(1-A)\exp\left(\frac{-i^2}{2\sigma_2}\right)\right],
    \label{eq:incs}
\end{equation}
where $A=0.95$, $\sigma_1=1.4^\circ$ and $\sigma_2=15.3^\circ$. This distribution is shown in figure~\ref{fig:inc_init}.

There is no preferred plane for the Companion to be captured on~\citep{Li2019}, which means that its expected inclination follows an isotropic probability distribution. Given the inclination distribution of the asteroids, for every Companion orbit there will be a subset of asteroids that have mutual inclinations greater than  $i_{\rm crit}$ that can undergo Kozai oscillations. A subset of these will reach maximum eccentricities such that their pericenters are interior to the orbit of jupiter. We can compute what fraction of asteroids reach this eccentricity by considering:
\begin{equation}
    e_{\rm max}\geq e_{\rm lim},
\end{equation}
where $e_{\rm lim}$ for each asteroid ($a$) is given by:
\begin{equation}
    (1-e_{\rm lim})a_{\rm a} \leq a_{\rm j},
\end{equation}
in which $a_{\rm j}$ is the semi-major axis of jupiter. Equation~\ref{eq:e_max} gives $e_{\rm max}$ for each asteroid with initial eccentricity $e_{0,a}$ and initial mutual inclination with respect to Companion $i_{0,a}$ as:
\begin{equation}
    \left(1-\sqrt{1-(5/3)(1-e_{0,a}^2)\cos^2 i_{0,a}}\right)a_{a}\leq a_{\rm j},
    \label{eq:ck}
\end{equation}
$i$ is given by:
\begin{equation}
    \cos i = \underline{\hat{J}}_{a}\cdot\underline{\hat{J}}_C,
\end{equation}
where $\underline{\hat{J}}_{a}$ and $\underline{\hat{J}}_C$ are the angular momentum unit vectors of each body. The angular momentum vector, in terms of the Keplerian angles; longitude of ascending node, $\Omega$, and inclination, $i$, is given as:
\begin{align}
    \underline{\hat{J}} &= \begin{bmatrix}
           \sin\Omega\sin i \\
           -\cos\Omega\sin i \\
           \cos i
         \end{bmatrix}.
\end{align}
With these limits in place we can integrate over the initial values of $a_{\rm a}$, $e_{\rm a}$, $\Omega_{\rm a}$ and $i_{\rm a}$ and determine the fraction of asteroids that get put on these kinds of orbits as a function of the inclination of the captured planet with respect to the reference plane.

\begin{figure}
\centering
\includegraphics[width = 1\columnwidth]{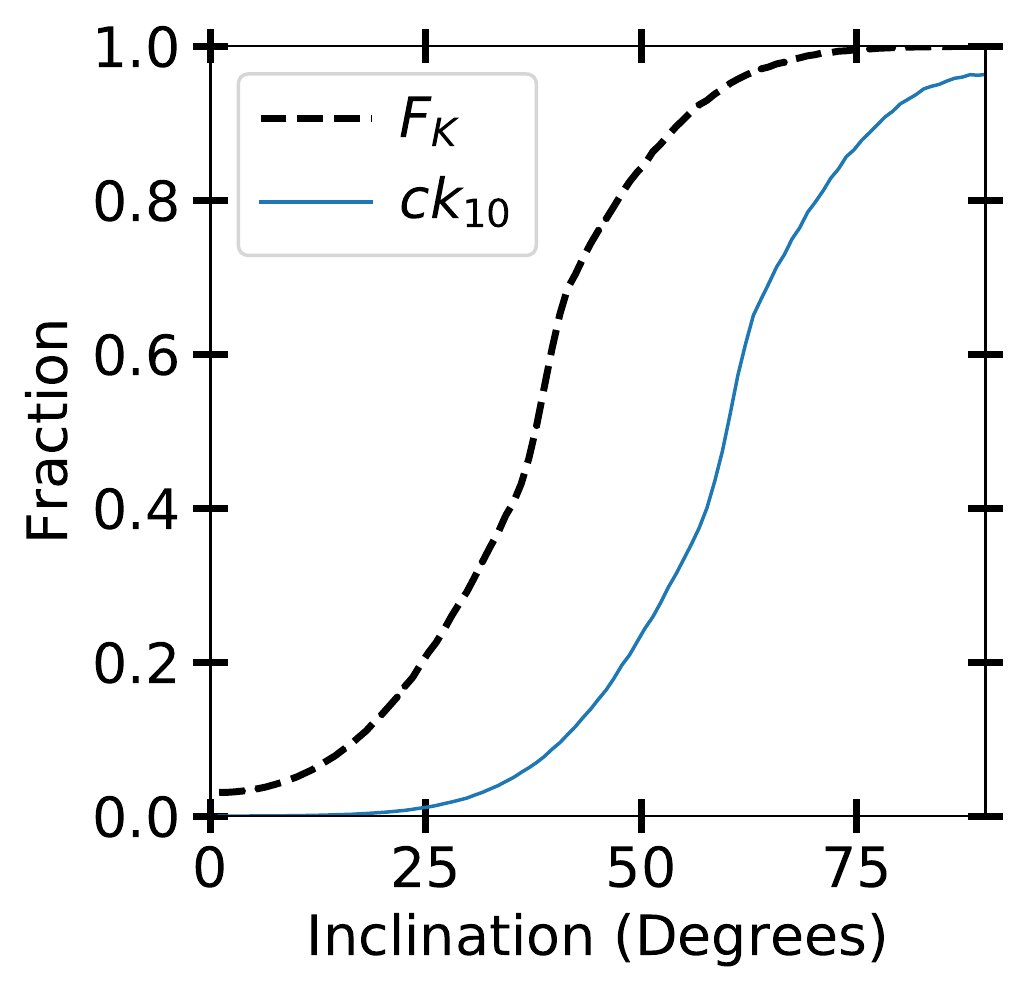}
\caption{The black dashed line shows the fraction of test particles that can undergo Kozai cycles as a function of the inclination of the captured giant. Whereas the blue line shows the fraction of test particles that achieve a pericentre less than 10 au, i.e. the test particles that satisfy equation~\ref{eq:ck}.}
\label{fig:initials}
\end{figure}

We show this fraction (blue line) for $a_{j}=10$ au in figure~\ref{fig:initials} along with the fraction of asteroids that can undergo Kozai cycles (black dashed line) for any given Companion inclination.

\subsection{SEjC}
We now go back to the SEjC system and simulate it with asteroids. The planetary system contains a one Solar mass star, the Earth, jupiter and a Companion. We set $a_{\rm j}$=10 au, $a_C$=100 au, $e_C$=0 and $m_\mathrm{j}= m_\mathrm{C}= {\rm M}_{\rm J}$. The inclinations simulated are $i_\mathrm{C}=[25,40,55,70,85]^\circ$. We place  Earth at 1 au and set it in the reference plane, whereas we give jupiter a small inclination (0-5 degrees) and a small eccentricity ($<$0.01). This means that the Earth will have exactly the same mutual inclination w.r.t. the Companion in every realization whereas for jupiter it will vary, mostly between $2-3$ degrees.

We do four different suites of runs, with three containing different subsets of asteroid trajectories as determined in figure~\ref{fig:initials}; the fourth being a retrograde version of the first:
\begin{itemize}
    \item full -- Set where the asteroid trajectories are simply drawn from the distributions described in the previous section -- 100 realizations per inclination with 200 asteroids each
    \item kozai -- Set where only such asteroids are drawn as to give Kozai oscillations, we do not do the 70 and 85 inclinations as figure~\ref{fig:initials} shows the fractions to be nearly identical to close kozai -- 100 realizations with 200 asteroids each
    \item close kozai -- Set where only such asteroids are drawn that give large enough Kozai oscillations to cross the orbit of jupiter -- 300 realizations per inclination, with 50 asteroids in the first 150 runs and 200 in the latter half of them.
    \item full retro -- Set where we reuse the initial conditions from the full simulations with only one thing changed, we set the companion on a ``mirrored" retrograde orbit. Given the broken symmetry we utilize figure~\ref{fig:kztest} and define a mirrored system such that a system is mirrored if the companion on a retrograde orbit gives the same $e_{\rm max}$ as its prograde equivalent. The mirrored inclinations are [25,40,55,70,85] $\rightarrow$ [161,146,136,125,-]. We do not do the 85 degree retrograde run as just as in the prograde run its mirror ends up pushing the planets into the star on a short time-scale, as seen in figure~\ref{fig:kztest} -- 100 realizations per inclination with 200 asteroids each.
\end{itemize}
We will refer to SEjC runs as a1.n where a $\in\{f, k, ck\}$ with f for full, k for kozai and ck for close kozai, and n$\in$\{1,5\} denotes the five different inclinations. All of the simulations run for $10^8$ years.

\section{Results}\label{sec:result}
\subsection{SEjC -- Asteroid fates}\label{sec:set1}
\begin{figure*}
    \centering
    \includegraphics[width=\textwidth]{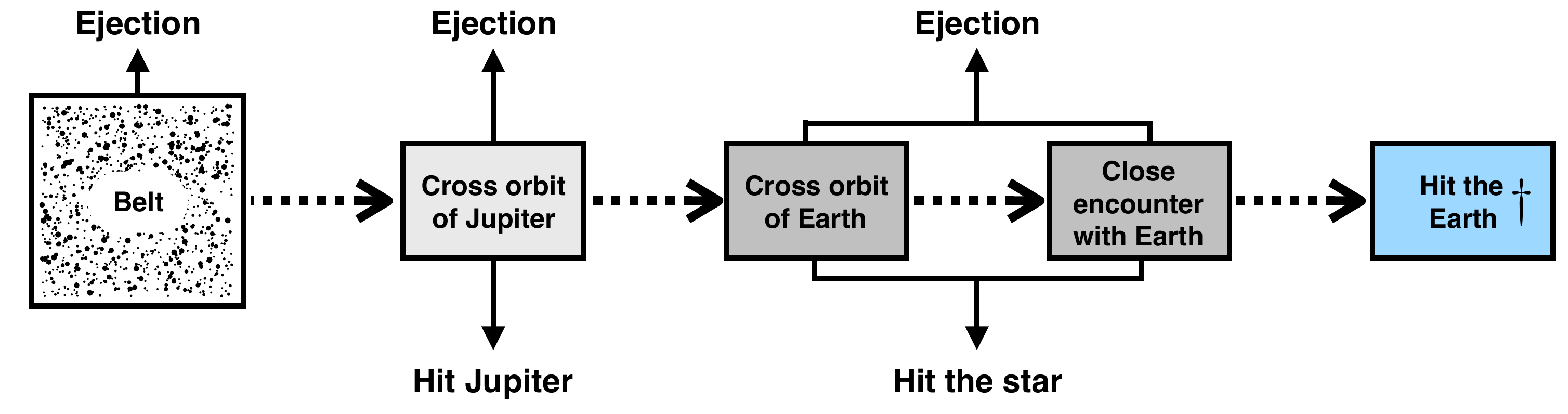}
    \caption{The steps an asteroid must go through in order to hit the earth and with the vertical lines showing the ways in which they can be removed from the simulation. In table~\ref{tab:encounter_table} we show two ratios: The number of asteroids that go through step 3 divided by the number going through step one, and the ratio of the numbers in step 4 and step 3. 
    Note: We will use the dagger throughout the paper with regards to hitting the Earth to remind the reader that we mean getting within 100 Earth radii.}
    \label{fig:flowchart}
\end{figure*}
\begin{figure}
    \centering
    \includegraphics[width = \columnwidth]{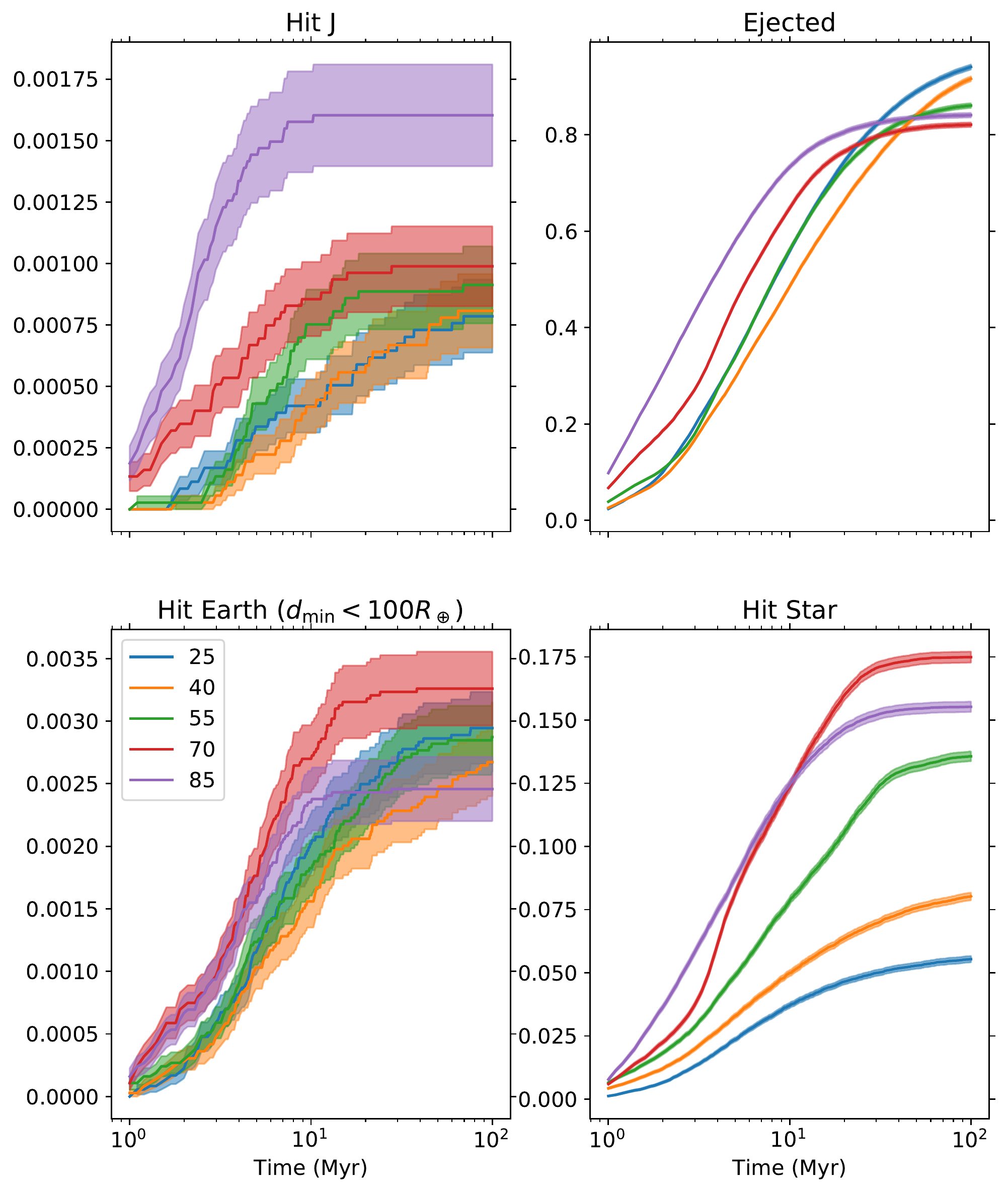}
    \caption{CDFs of the four different asteroid removal mechanisms in the ck runs. Each color represents a different inclination and the shaded area shows their respective one sigma error-bar. }
    \label{fig:fates}
\end{figure}

Before even considering what happens with the asteroids we consider jupiter as it too will undergo Kozai cycles, although it should be noted that the precession from a second giant planet in the system would have blocked the cycles. We can see in figure~\ref{fig:kztest} that jupiter reaches high enough eccentricities ($>0.9$) to cross the orbit of the Earth for companion inclinations between 79 and 117 degrees. This means that in 32.4\% of cases jupiter will reach sufficient eccentricities to cross the orbit of Earth leading to either a collision or a scattering which can either eject the Earth or send it into the Sun. We do indeed see this in the ck1.5 and f1.5 simulations but we report the behaviour of the asteroids from those simulations regardless as most of the asteroids will undergo several Kozai cycles and have time to reach their end state long before the 20-30 Myr it takes for the jupiter to increase its eccentricity and encounter the Earth.

We now consider what happens to the asteroids that are removed from the asteroid belt. Figure~\ref{fig:flowchart} shows a flowchart of what the asteroids do before they hit the Earth${}^\dagger$. From left to right we see the steps each asteroid must go through before they hit the Earth with the perpendicular arrows showing how the asteroids can be removed along the way. The flow is as follows: All the asteroids start in the belt, some are then made to cross the orbit of jupiter either purely through Kozai oscillations or through Kozai and additional interactions, a subset of these asteroids end up crossing the orbit of the Earth at which point they can have close encounters with the Earth. Roughly 10\% of the bodies that have close encounters end up hitting the Earth${}^\dagger$.

There are four ways in which the asteroids are removed from the simulations: ejection, getting within 0.2 au of the star and hitting jupiter or Earth${}^\dagger$. We can see how the likelihood of each end state varies with inclination in the ck runs as they have been selected to cross the orbit of jupiter. We show this in figure~\ref{fig:fates} by looking at CDFs, along with their one sigma error-bars. The biggest differences are seen in the number of asteroids that are removed by ejection and the number removed by getting too close to the star. At high inclinations we see a much larger fraction hitting the star, this is because at higher inclinations the Kozai cycles are faster, allowing asteroids to more easily reach sufficiently high eccentricities to be removed by the star before encountering jupiter. The reason we see more asteroids hitting the star in ck1.4 vs ck1.5 is that  jupiter gets to very high eccentricities much faster in ck1.5 allowing it to clear out more asteroids before they are all gone. The spike seen in the jupiter-hits for ck1.5 is due to asteroids having smaller pericentres when encountering jupiter, giving higher $\Delta V$ making the encounters more likely to result in a collision.

\subsection{SEjC -- Earth impacts}

\begin{figure}
\centering
\includegraphics[width = 1\columnwidth]{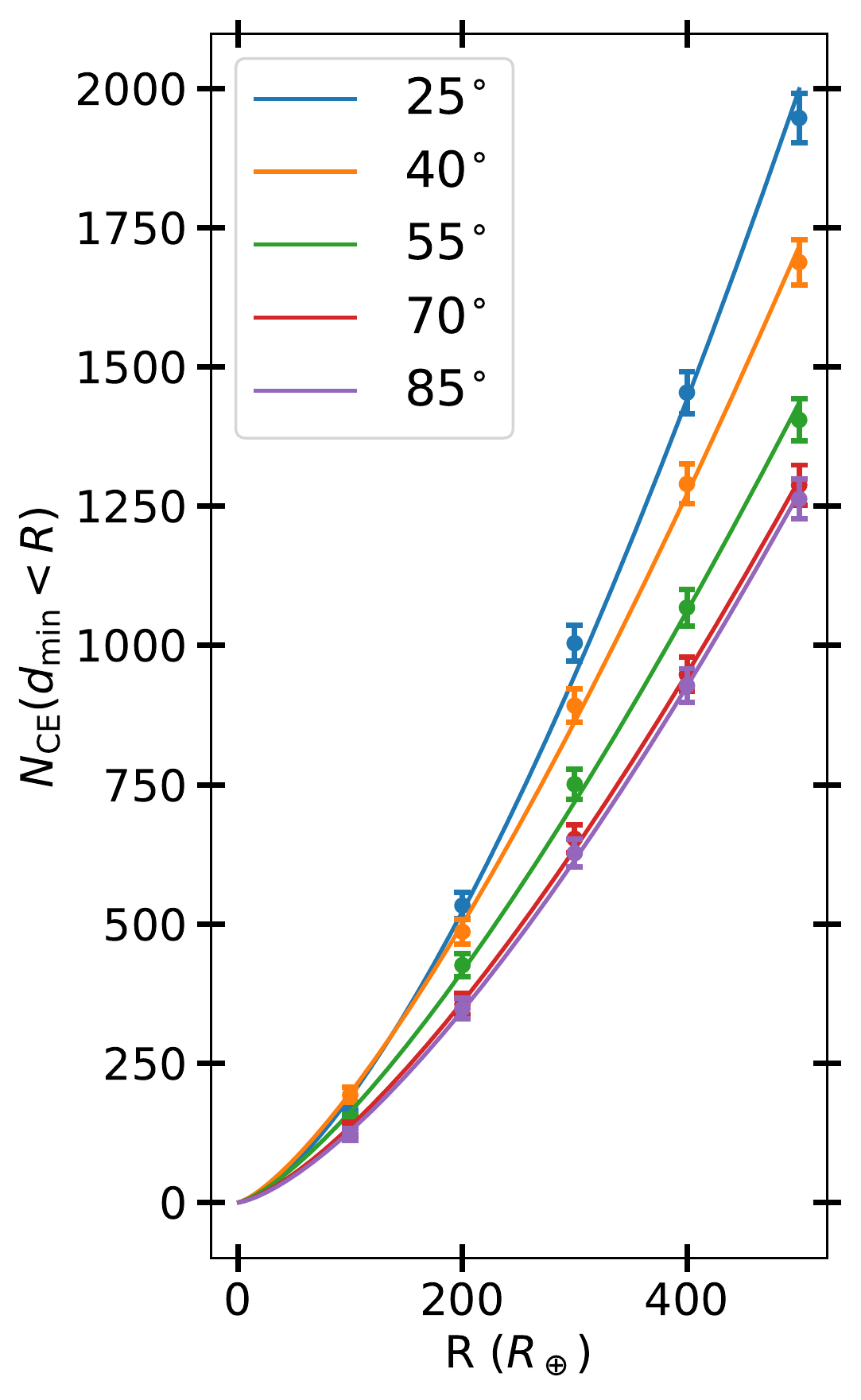}
\caption{The number of asteroids that have an Earth--asteroid close encounters where the closest approach is less than radius R for the five different inclinations.}
\label{fig:fit}
\end{figure}

To determine the number of asteroids that would actually hit the Earth we look at the number of close encounters the inflated Earth has, which are logged between 100 and 500 Earth radii. We utilize the ck runs as they have the largest number of asteroids having close encounters with the Earth. For each ck run we take the number of close encounters as a cumulative function of the closest approach (in terms of Earth radii) and fit a power law to it, as shown below:
\begin{equation}
    F = a\times R^b.
    \label{eq:power}
\end{equation}
We show the data and the fits for the ck runs in in figure~\ref{fig:fit} with the values of a and b shown in table~\ref{tab:ab}.\\
\begin{table}
    \centering
    \caption{The parameter values for the fits to the close encounters using equation~\ref{eq:power} on the ck runs}
    \begin{tabular}{c|c c}
    Inclination & $a$ & $ b$\\
    \hline
    25 & $0.21\pm0.06$ & $1.36\pm0.05$\\
    40 & $0.15\pm0.04$ & $1.42\pm0.05$\\
    55 & $0.13\pm0.04$ & $1.46\pm0.05$\\
    70 & $0.19\pm0.05$ & $1.43\pm0.04$\\
    85 & $0.10\pm0.03$ & $1.51\pm0.05$\\
    \hline
    \end{tabular}
    \label{tab:ab}
\end{table}

\begin{figure}
    \centering
    \includegraphics[width=\columnwidth]{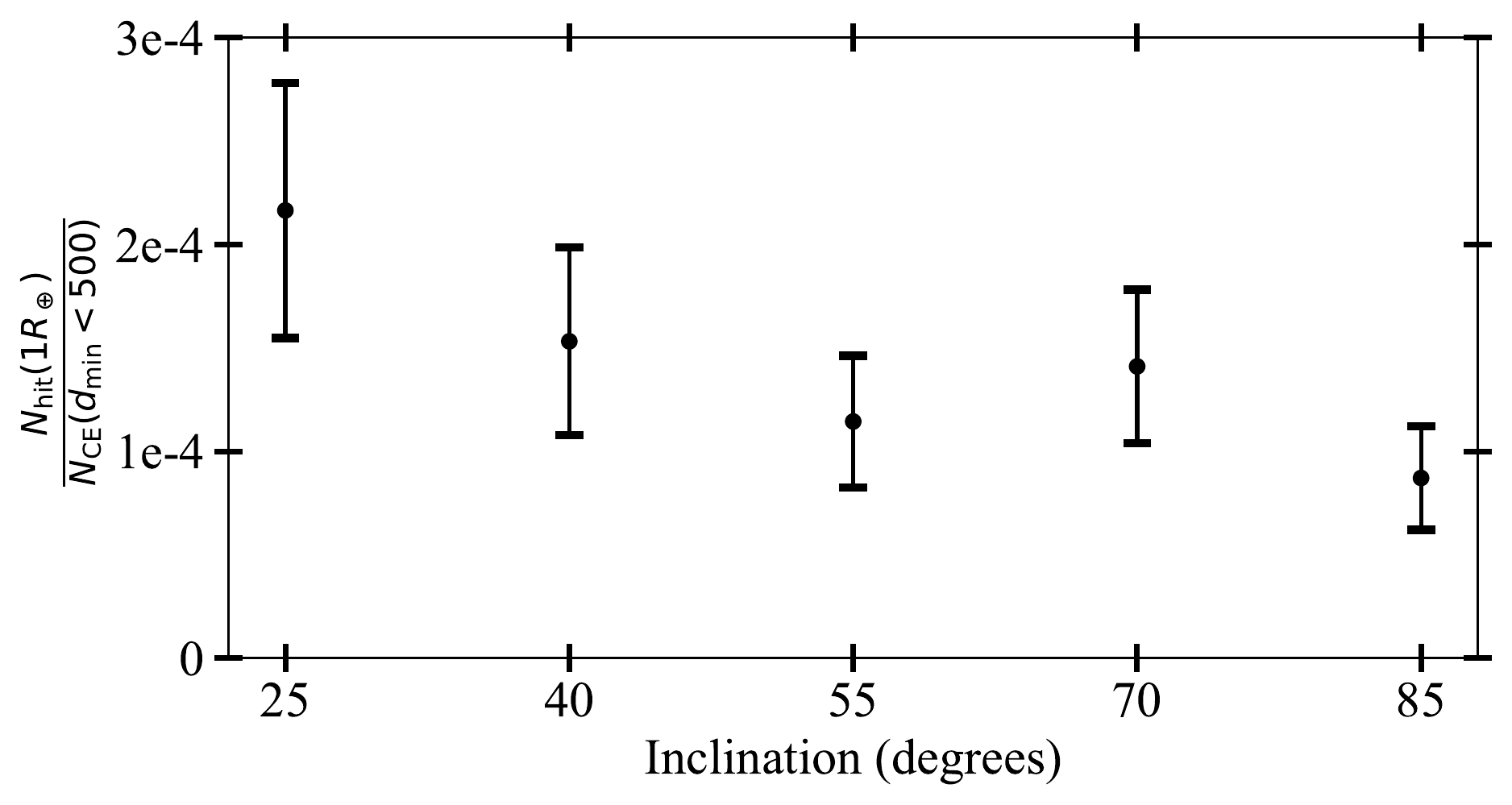}
    \caption{The number of asteroids expected to hit an Earth-sized planet in our simulations divided by the number that get within 500 Earth radii of the Earth.}
    \label{fig:norm_error}
\end{figure}

To compare the outcomes of the runs at the different inclinations we compare the ratio between the number of expected hits for an Earth-sized body with the number of asteroids that get within 500 Earth radii. We plot these ratios for the different runs in figure~\ref{fig:norm_error}. We see no clear trend with inclination and conclude that the outcomes at the different inclinations are consistent with each other.

This means that to get the fraction of asteroids in the ck runs that would hit an Earth-sized planet we can stack the data from all the ck runs and compute the fraction. We find the fraction to be $(4.0\pm0.5)\times10^{-6}$. This fraction is interpreted as the fraction of asteroids removed from the asteroid belt that end up hitting the Earth. The number of impacts this corresponds to depends on what fraction of the asteroid belt a given Companion destabilises and the number of bodies in the asteroid belt, this is discussed in more detail in section~\ref{sec:disc}.

\subsection{SEjC -- The effect of jupiter}

The ck selection gives asteroids on very different initial orbits, e.g. $i_C=25^\circ$ gives only the asteroids with largest initial inclination to the reference plane whereas an $i_C=85^\circ$ will essentially select asteroid orbits randomly. The lack of correlation between the fraction of asteroids hitting the Earth with $i_C$ implies that the asteroids that have a close encounter with jupiter and do not get ejected have their orbit changed to such a degree that their previous orbits do not matter.

This is further exemplified in table~\ref{tab:encounter_table} where we show two fractions from each system and run. $f_{\rm ce}$ is the fraction of asteroids that cross the orbit of jupiter that get within 500 Earth radii of the Earth and $f_{\rm hit}$ is the fraction of those that get within 500 Earth radii that end up hitting the Earth${}^\dagger$. We see in table~\ref{tab:encounter_table} that $\sim$3\% of asteroids crossing the orbit of jupiter have a close encounter with the Earth and that out of those $\sim$10\% end up hitting the Earth${}^\dagger$. This means that 3 out of 1000 asteroids crossing the orbit of jupiter will hit the Earth${}^\dagger$ regardless of both their initial orbit and the Companion's orbit.

\begin{table}
\caption{$f_{\rm ce}$ and $f_{\rm hit}$ are two fractions, showing the fraction of asteroids that cross the orbit of  jupiter that get within 500 Earth radii of the Earth and the fraction of those that get within 500 Earth radii that end up hitting the Earth${}^\dagger$.\newline
* Indicates runs where the outcome has $<$50 asteroids.}
    \centering
    \begin{tabular}{c| l l }
    System & $f_{\rm ce}$ & $f_{\rm hit}$\\
    \hline
    ck1.1 & 0.028 & 0.106 \\ 
    ck1.2 & 0.028 & 0.099 \\ 
    ck1.3 & 0.032 & 0.092 \\ 
    ck1.4 & 0.035 & 0.097 \\ 
    ck1.5 & 0.031 & 0.083 \\ 
    k1.1 & 0.026 & 0.108*\\
    k1.2 & 0.032 & 0.088*\\ 
    k1.3 & 0.034 & 0.080*\\ 
    f1.1 & 0.030*  & 0.034* \\ 
    f1.2 & 0.024 & 0.107*\\
    f1.2r & 0.019* & 0.093*\\
    f1.3 & 0.031 & 0.062*\\ 
    f1.3r & 0.031 & 0.076*\\
    f1.4 & 0.036 & 0.068*\\
    f1.4r & 0.036 & 0.064*\\
    f1.5 & 0.034 & 0.093\\ 
    \hline
    \end{tabular}
    \label{tab:encounter_table}
\end{table}

\subsection{SEjC - Asteroid removal from belt}

In table~\ref{tab:Time_limitied_fracs} we show the outcome of the asteroids in the k and f runs at two different times, 10 and 100 Myrs. We pick 10 Myr because during this time the asteroids undergo multiple cycles whereas jupiter does not have time to increase its eccentricity. If only the subset of asteroids that are simulated in the ck runs end up crossing the orbit of jupiter we would expect the fractions crossing, $f_{\rm cross}=ck_{10}$ for the full runs and $f_{\rm cross}=ck_{10}/F_K$ for the k-runs. The table show that afyer 10 Myr, the fraction of asteroids removed is mostly consistent with the number of asteroids that can cross the orbit of jupiter, whereas at later times significantly more of the asteroids have been removed.

This means that there are two processes removing asteroids from the system. The first and faster one is jupiter removing the asteroids that have their eccentricity pumped up by the Companion. The second process is a combination of asteroids being pushed to higher eccentricities through interactions with the Companion and jupiter eventually putting them on a jupiter-crossing orbit leading to their eventual removal, and asteroids being removed directly by the Companion from the belt. The latter of which can happen because the Companion encounters the asteroids at apocentre giving encounters with very low $\Delta V$ during which the Companion can do enough work to unbind the asteroids.

\subsection{Examining longer timescales}

\begin{figure}
    \centering
    \includegraphics[width = \columnwidth]{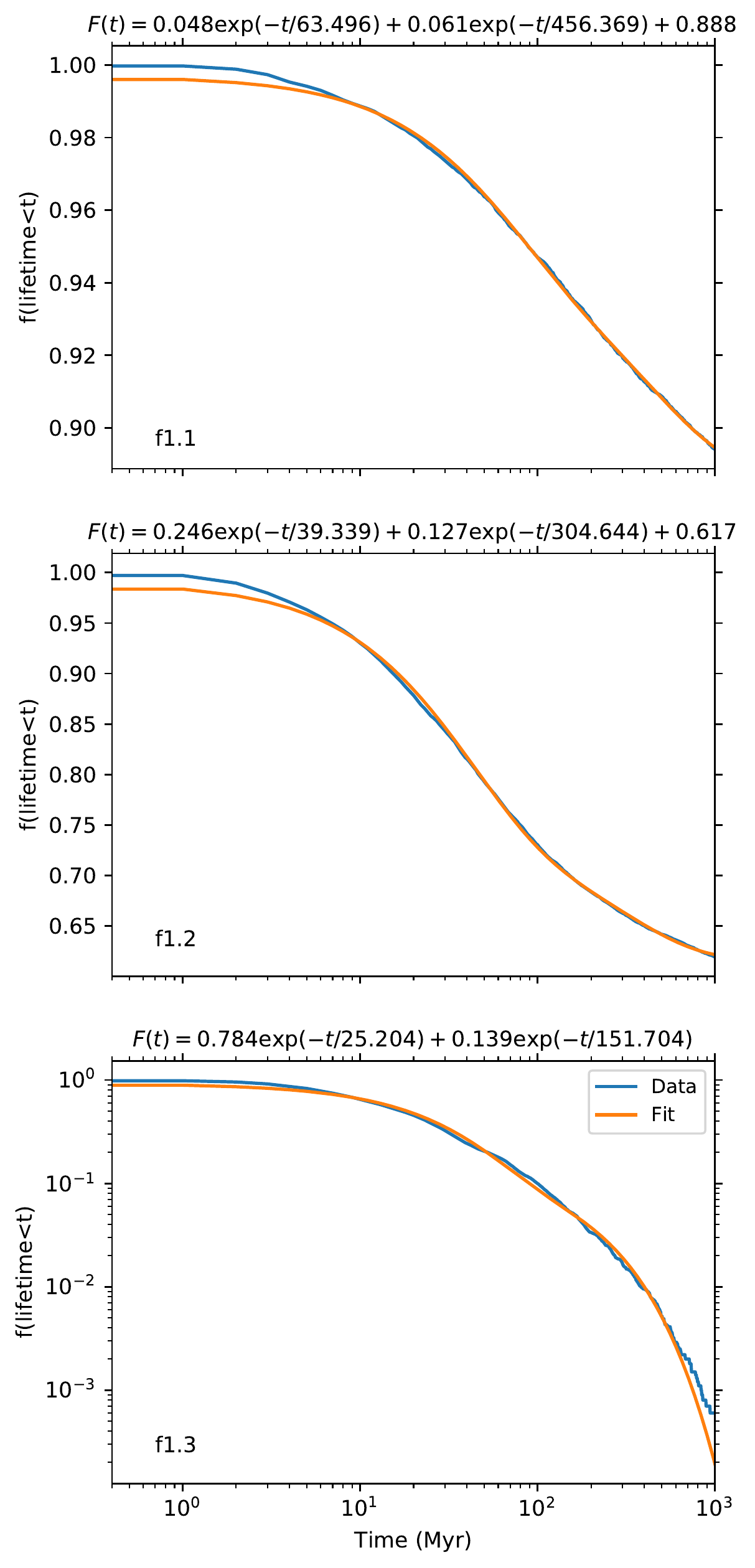}
    \caption{The fraction of asteroids remaining as a function of time for three different 1 Gyr runs with the full asteroid belt. The blue line shows the data with the best fit plotted on top of it. The values for each of the best fits is shown at the top of each panel. \newline Note: the y-axis changes scale (and from linear to log) between the panels}.
    \label{fig:depletion}
\end{figure}

We do an extra set of 1 Gyr simulations, in which we remove the Earth for computational efficiency, for runs f1.1, f1.2 and f1.3 to see what fraction of test particles end up being removed. Given the two processes described in the previous section we fit a double exponential to the fraction of asteroids remaining in the asteroid belt as a function of time, on the form shown below:
\begin{equation}
    F(t)=a\times\exp\left(-\frac{t}{b}\right)+c\times\exp\left(-\frac{t}{d}\right)+g
    \label{eq:time_fit}
\end{equation}
The three rates with their respective best fits are shown in figure~\ref{fig:depletion}. We see that the depletion is modeled well with the double exponential and that as expected there is one fast process and one slower process.

However, the faster decaying exponential removes a larger fraction of the asteroids than the ck subset. We take this to mean that each exponential captures the removal done by each body, rather than the aforementioned removal processes.

\begin{table}
    
    \caption{The outcomes of the kozai and full runs. $f_{\rm expected}$ is the fraction we expect to cross the orbit of J if only the ck subset were the ones crossing. $f_{\rm cross,10}$ and $f_{\rm cross,100}$ are the fractions of asteroids that end up crossing the orbit of jupiter after 10 and 100 Myr. $f_{\rm removed}$ is the fraction of asteroids removed after 100 Myr, either by being ejected or hitting the star or planets.\newline
    Note: the output has a resolution in time of 10 kyr, i.e. if a particle crosses 10 au and is ejected within 10 kyr it will never be logged as having crossed the orbit of jupiter, resulting in an underestimation of $f_{\rm cross}$.}
    \begin{center}
    \begin{tabular}{c| c | c c c}
    System & $f_{\rm expected}$ & $f_{\rm cross,10}$ & $f_{\rm cross,100}$ & $f_{\rm removed}$  \\
    \hline
    k1.1 & 0.05 & 0.06 & 0.20 & 0.24\\
    k1.2 & 0.13 & 0.11 & 0.30 & 0.41\\
    k1.3 & 0.36 & 0.36 & 0.77 & 0.94\\
    f1.1 & 0.01 & 0.01 & 0.05 & 0.05\\
    f1.2 & 0.09 & 0.08 & 0.23 & 0.27\\
    f1.2r & 0.04 & 0.04 & 0.11 & 0.14\\
    f1.3 & 0.33 & 0.39 & 0.83 & 0.91\\
    f1.3r & 0.14 & 0.12 & 0.47 & 0.54\\
    f1.4 & 0.79 & 0.81 & 0.95 & 1.0\\
    f1.4r & 0.33 & 0.38 & 0.84 & 0.95 \\
    f1.5 & 0.95 & 0.91 & 0.95 & 1.0\\
    \hline
    \end{tabular}
    \end{center}
    \label{tab:Time_limitied_fracs}
\end{table}

\section{Discussion}\label{sec:disc}

\subsection{Initial impact rates}
\begin{figure}
    \centering
    \includegraphics[width=\columnwidth]{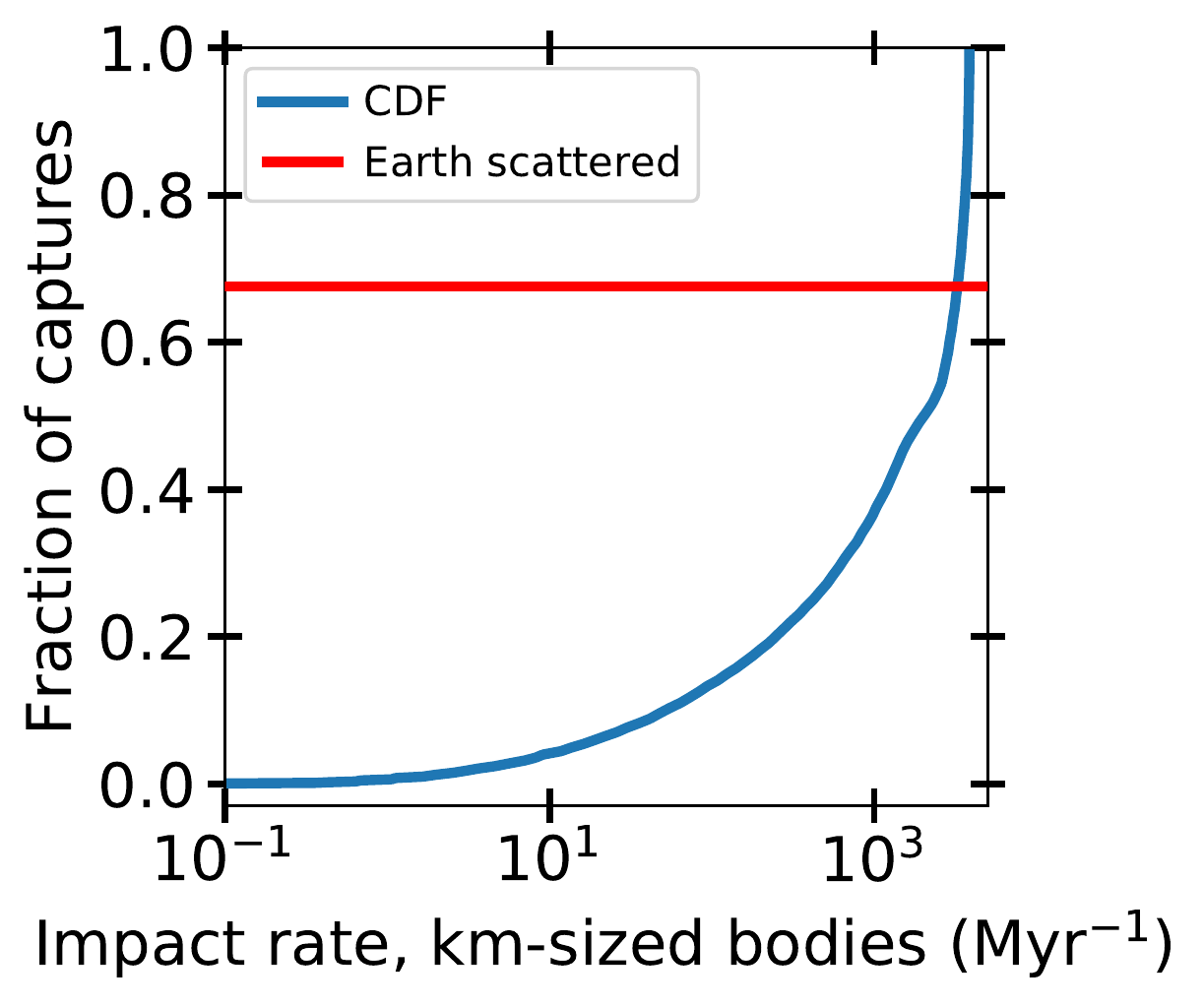}
    \caption{The blue line shows the CDF of the average number of impact per Myr of bodies greater than 1 km as a function in the first 10 Myr following the capture of a planet on isotropically distributed orbit orientations. The numbers are calculated assuming a $0.01M_\oplus$  Kuiper belt. The red line shows the fraction of outcomes in which the Earth either ends up being ejected or scattered and removed from the system.}
    \label{fig:impacts}
\end{figure}

We showed in section~\ref{sec:set1} that 4 in $10^6$ jupiter-crossing asteroids will impact the Earth. We see from table~\ref{tab:Time_limitied_fracs} that in the first 10 Myr the number of jupiter-crossing asteroids is well approximated by the ck-fraction shown as a function of Companion inclination in figure~\ref{fig:initials}. From \cite{Schlichting2012} we can compute the number of $>1$ km sized bodies in the present day Kuiper Belt to be $\sim10^{10}$ with models for the early evolution of the Solar System giving a primordial Kuiper belt 2000-4000 times more massive (see e.g.~\citealp{Gomes2005, Gomes2018}). Given these three things we can compute the average impact rate during the first 10 Myrs following the capture of the Companion. This is shown in figure~\ref{fig:impacts}, where we plot the CDF of the average impact rate distribution for $>1$ km sized bodies using the present day Kuiper belt. The red line in the figure shows that roughly one third of captures will result in jupiter eventually ejecting the Earth, this however happens on a longer time-scale meaning that we can determine the impact rate before the orbits of the planets cross and for that reason we can include rates above the red line in the figure.

We can compare the rates we find with the inferred present-day impact rate on the Earth.. It should however be noted that the vast majority of Earth-impactors are asteroids originating in the asteroid belt and not the Kuiper belt~\citep{Strom2005}. Whilst not very well constrained, the most most recent estimates give a rate of $\sim$1 per Myr for asteroids with a radius of 1 km~\citep{Wheeler2019}. The average rate following a capture, shown in figure~\ref{fig:impacts}, is obviously significantly higher than 1 per Myr. Such a high impact rate would be quite harmful for life developing on Earth, as it also implies a higher impact rate for larger asteroids such as the one that wiped out the dinosaurs. And it would be even more harmful for any developing civilisation as even a km sized body can cause a period of global cooling~\citep{Toon2016} since long term climate stability is critical for the development of agriculture~\citep{Bettinger2009}.

However, given that planets are pretty much exclusively captured in the birth clusters of stars the effect of this initial very high impact rate is unlikely to affect any planet hosting life. When we do consider the Sun in its birth cluster with a primordial Kuiper Belt of $20-40M_\oplus$, the impact rate following the capture of a planet then would be in line with the rate observed for the Late Heavy Bombardment~\citep{Zahnle2003}.

\subsection{Later impact rates}

Here we consider the impact rates at later times, rather than the initial surge discussed previously. For this we look at the 1 Gyr runs of f.1-3 without the Earth. We can determine the removal rate of asteroids by differentiating the equations shown in figure~\ref{fig:depletion} and estimating that out of the ones that are removed from the asteroid belt 4 in $10^6$ hit the Earth.

1 Gyr after the capture, for the three different inclinations and a $0.01M_\oplus$ Kuiper belt, we find impact rates of $>1$ km sized objects per Myr of: [0.60, 0.62, 0.051]. And, it takes [764, 858, 547] Myr for the flux to reach the current flux on Earth of 1 per Myr. This means that a capture on a low inclination might actually be worse for Earth than a high one (assuming it is not so high as to destabilize it). This is because, rather than a short cataclysmic period during which most of the Kuiper belt is depleted, it instead results in an increased rate over a much longer time, giving life on the planet less time to recover and develop between potential extinction events.

\subsection{The companion as a transient perturber}

Concerning our Solar system, using wide field infra-red data from WISE, \cite{Luhman2014} excludes the possibility of having any planet above the mass of Saturn within $28\,000$ au and mass of Jupiter within $82\,000$ au, excluding the possibility of the Sun having a Companion such as the one we have simulated. However, as we have shown, many companions can deplete a large fraction of the the Kuiper-belt in only a few  Myr. This means that most of the work could have been done whilst the Sun was still in its birth cluster. Therefore, even if the Companion is removed in a subsequent encounter a couple of Myr after capture it can already have removed a large fraction of the Kuiper Belt. In fact,~\cite{Malmberg2007} finds that stars that have a close encounter in their birth cluster, have a higher probability of having another close encounter than they initially had to have one at all. Given that a captured Companion that affects the Kuiper belt but does not leave a significant dynamical imprint on the inner planets must be on quite a wide orbit, it can be unbound rather easily in subsequent close encounters in the birth cluster. If the Companion is not removed in the birth cluster it cannot later be removed by other stars as sufficiently close encounters only occur once every $10^{11}$ years or so~\citep{Bailer-Jones2018}, but it could have been removed by a giant molecular cloud. From~\cite{Gustafsson2016} we see that the tidal field of a massive GMC ($\sim10^6M_\odot$) could do enough work on the planet to unbind it as the star passes through the GMC. Such encounters would occur approximately once per Gyr~\citep{Kokaia2019a}.

\subsection{Water delivery}

It has been established by looking at isotope ratios of the Earths' water that a significant fraction of it was delivered from the outer Solar system (see e.g.~\citealp{Wu2018}), which is in line with planet formations models because Earth being inside of the water-ice line makes it very unlikely that the water on the surface of the Earth condensed as the Earth was forming. The Earth's surface holds approximately $10^{21}$ kg of water, or 0.00023$M_\oplus$. This corresponds to $\sim5\times10^{-6}$ of the primordial Kuiper belt or $\sim10^8$ kilometer-sized asteroids made completely of water-ice. 

With jupiter on its own having a throughput of $4\pm0.5\times10^{-6}$ in the SEjC simulations, it seems unlikely that with the addition of Saturn in our Solar system that enough of the Kuiper belt would make it through to deliver a significant fraction of the water on Earth. Which also means that the water must have been delivered before most of the debris was removed from the Solar system as otherwise not enough of it could have made it past the orbits of Jupiter and Saturn.

\subsection{Generalizing farther}
\begin{figure}
    \centering
    \includegraphics[width= \columnwidth]{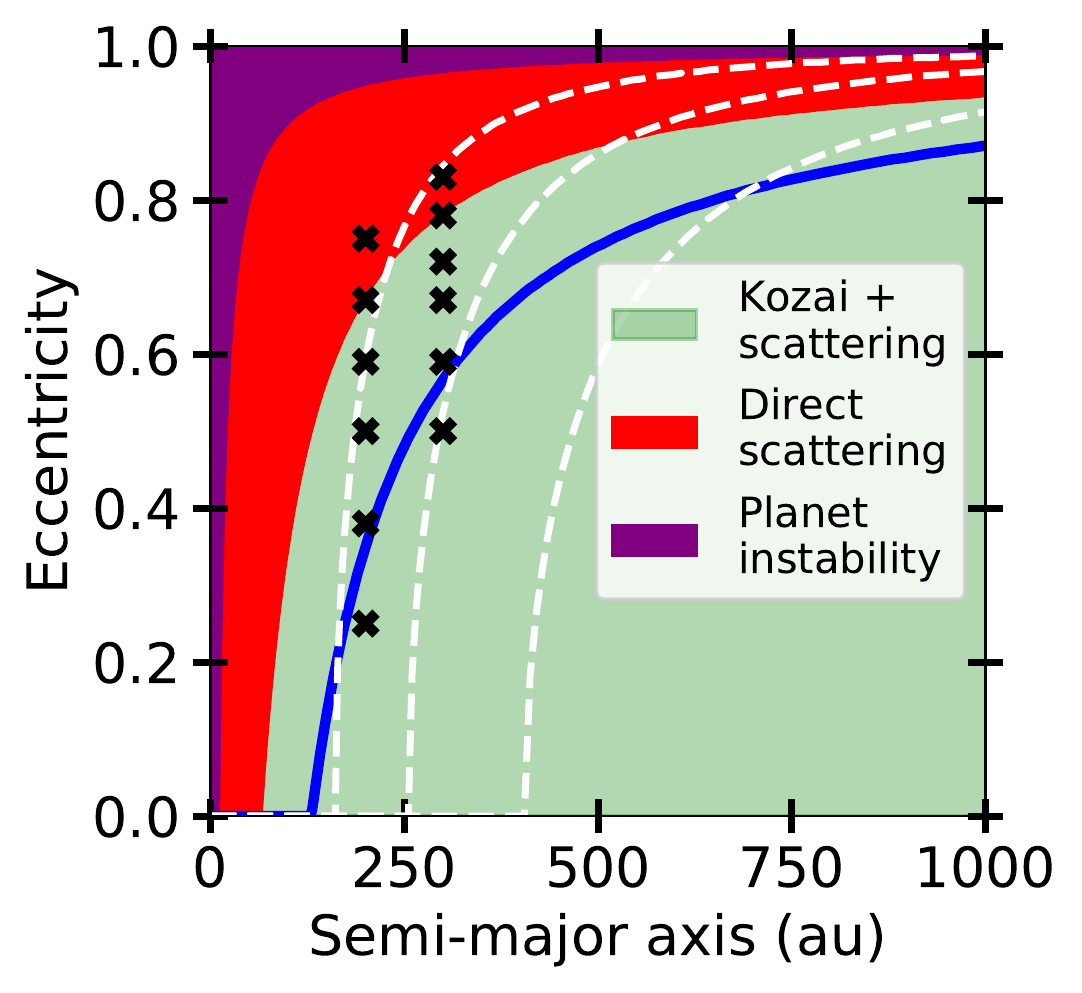}
    \caption{The different effects that are dominant for different $a$ and $e$ for captured planets. Planets captured in the purple region will disrupt the whole system. In the red region they will be orbit-crossing with the asteroid belt and disrupt it. In the green region asteroids need to have their eccentricities increased through Kozai oscillations in order to be scattered by either the planets in the system or the companion, where the blue line is a line of constant pericenter within which the companion can scatter the asteroids. The black crosses show where we have done 1 Gyr long simulations with a Jupiter-mass companion and a Jupiter-mass internal planet at 5 au in order to determine what fraction is depleted. The white lines show where the precession from a Jupiter-mass inner planet with a semi-major axis of (from left to right) 10, 5 and 2.5 au blocks the Kozai cycles of the asteroids.}
    \label{fig:regimes}
\end{figure}

We have interpreted that the depletion of the asteroid belt is driven by both the orbit of the jupiter as well as the companion as the depletion is well modelled by a two-component exponential. These processes are however not decoupled from each other as the interaction between belt and jupiter is driven by the Kozai cycles which the Companion induces in the belt, the time-scale and magnitude of which is set by the Companions' orbit as shown in eq.~\ref{eq:Kozai}. The orbit of the Companion can therefore set three different depletion regimes for the asteroid belt (four if one considers planet stability separately):
\begin{enumerate}
    \item The Companion can orbit-cross the asteroid belt disrupting it directly.
    \item The Companion can pump up the eccentricities of the asteroids with either the Companion itself or jupiter ejecting them later.
    \item The Companion can pump up the eccentricities of the asteroids and then have them be removed by only jupiter.
\end{enumerate}
Given these differences it becomes impossible to extend our initial semi-analytic analysis farther. To explore the outcome of companions captured on their possible different orbits we do a suite of low resolution (1000 asteroids per run) simulations. These simulations are shown in figure~\ref{fig:regimes} as black crosses with the different regimes highlighted. In these simulations we move jupiter to 5 au as it increases the precession time by a factor of 4 (equation~\ref{eq:Ptimescale}) and increases the region in which the captured Companion gives Kozai oscillations; as shown by the white dashed line in figure~\ref{fig:regimes}. The semi-major axes and eccentricities are chosen such that there is a constant pericentre, with pericentre distances of: 50, 65, 83, 100, 125 and 150 au. For each black cross we simulate 6 different companion inclinations, 40, 55 and 70 degrees prograde and also 34, 44 and 55 degrees retrograde, which are the inclinations selected by looking at figure~\ref{fig:kztest} as described in section~\ref{sec:setup}. After performing the simulations we once again fit equation~\ref{eq:time_fit} to the depletion and differentiate it to get the removal rate, then working under the assumption that the throughput as depicted in figure~\ref{fig:flowchart} remains the same we can also compute the impact rates.

Figure~\ref{fig:deptime} shows for how long each run results in an impact rate higher than that of the Earth. In it we see that the runs in the red regime in figure~\ref{fig:regimes} give a very short enhancement as the whole belt is disrupted very quickly. As the Companion's pericentre moves outwards we see the enhancement time increasing because the depletion rate goes down. This is in part because of the increasing Kozai time-scale but more than that it is that the fraction of asteroids that \textit{can} have their orbits disrupted goes down as the Companion's pericentre increases due to moving from regime 2 to 3 which is confirmed by the decreasing number of close encounters for the Companion. We conclude that the scenario described in figure~\ref{fig:regimes} is largely correct, but that the transitions are not as sharp as shown because the asteroids' semi-major axes range between [40,50] so the correct way to view it is to say that each asteroid is in a specific regime with respect to the Companion.
\subsubsection{Mapping the outcome-space}

\begin{figure}
    \centering
    \includegraphics[width = 1\columnwidth]{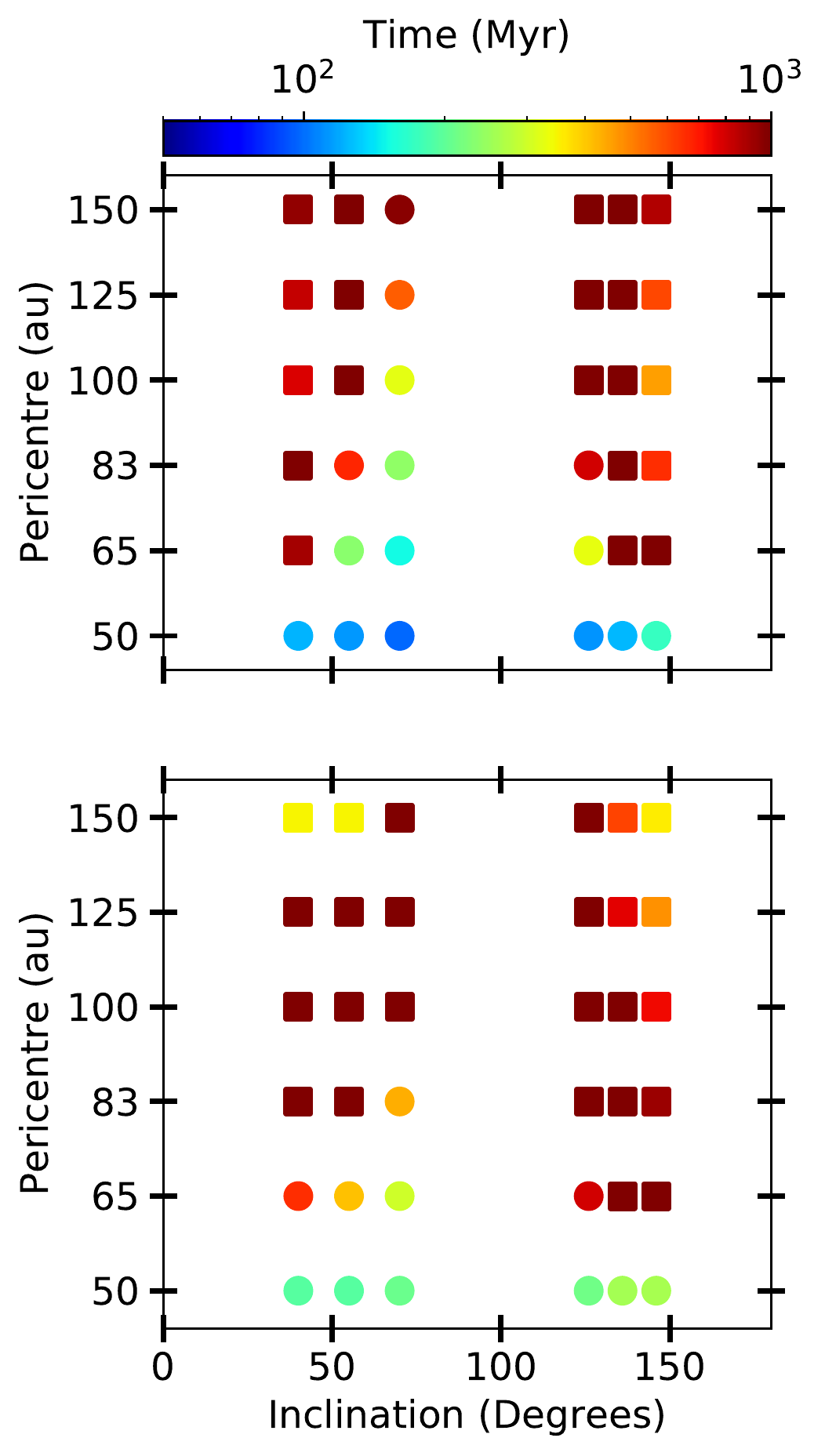}
    \caption{The figure shows for how long a captured companion produces a higher impact rate of $>1$km sized objects compared to that of the present day Earth, for each cross in figure~\ref{fig:regimes}. For each of those crosses 6 simulations at different inclinations have been performed as shown by the markers in each panel. The square marker indicates that there were asteroids remaining at the end of the run, whereas the circles show runs where all the asteroids were removed. Marker color shows the time until the Earth rate is reached after capture, with 1 Gyr set as an upper limit.  \newline
    Top panel shows runs with $a=200$ au and bottom shows runs with $a=300$ au.}
    \label{fig:deptime}
\end{figure}

For each run we also log the mean enhancement in impact rate and use this to map the outcome-space following Companion capture. We explore the outcome-space by drawing Companions with random eccentricities and inclinations at the two semi-major axes. The eccentricity is drawn from a thermal distribution and the inclination from an isotropic distribution~\citep{Li2019}. For each of the $40\,000$ Companions drawn we interpolate between the points in figure~\ref{fig:deptime} and determine the time during which the impact rate is enhanced to be greater than the impact rate on present Earth and also the mean impact rate during this time-period.

\begin{figure}
    \centering
    \includegraphics[width = 1\columnwidth]{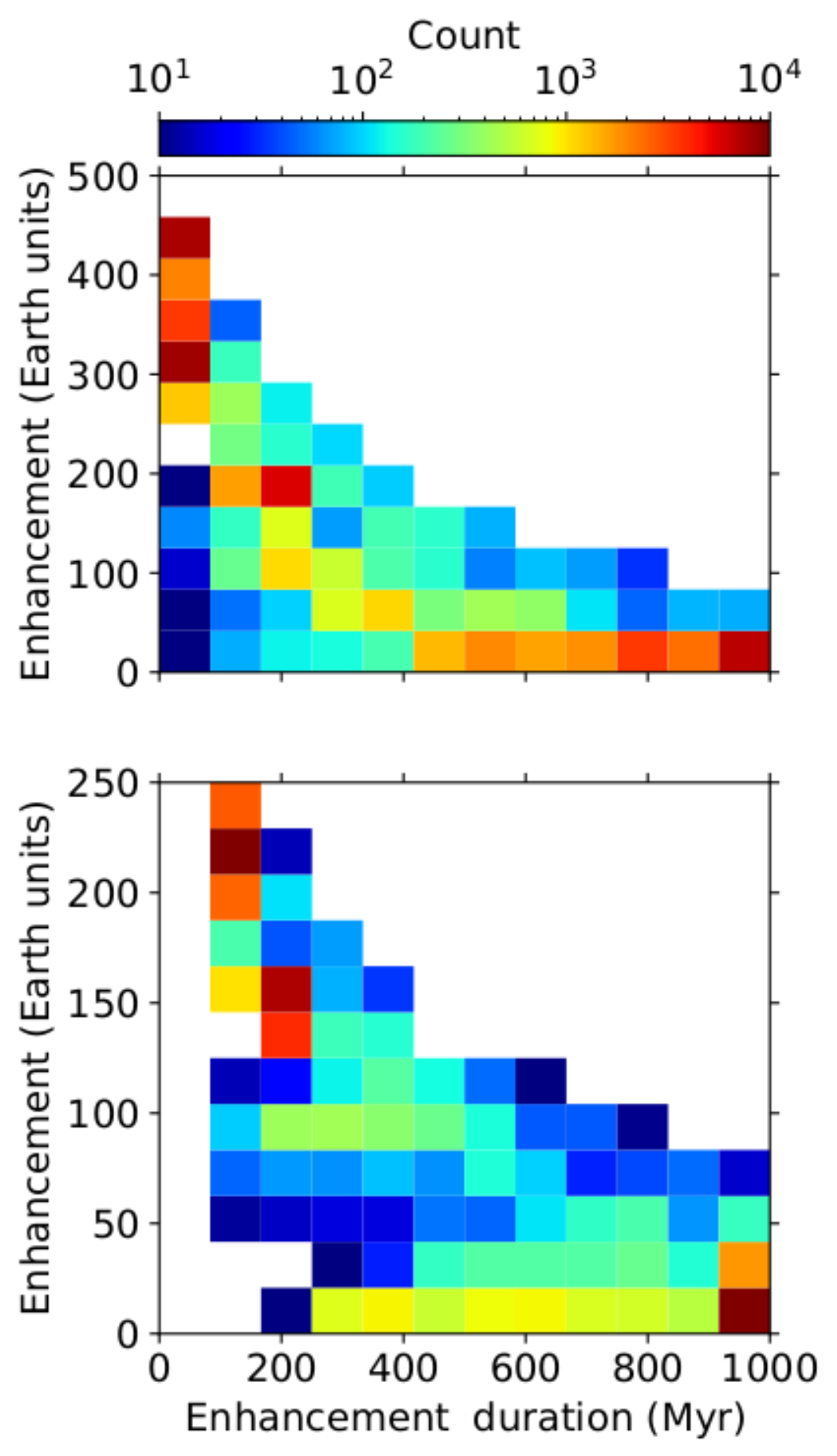}
    \caption{The outcome space of enhancement duration and corresponding mean enhancement for captured companions with semi-major axis of 200 (top) and 300 (bottom). The color shows what number out of the $40\,000$ random Companions drawn give an outcome in each bin. The figure is generated by drawing  eccentricities from a thermal distribution and inclinations from an isotropic distribution and then interpolating in figure~\ref{fig:deptime}.}
    \label{fig:heatmap}
\end{figure}

The results of this is shown in figure~\ref{fig:heatmap}. The first thing to note is that each panel only shows approximately 60\% of the outcomes. At both semi-major axes nearly 40\% ($\sim$30\% due to Companion inclination and $\sim$10\% due to capture on high eccentricity) of captured Companions lead to the system becoming unstable.

For the remaining $\sim$60\% the outcome is more or less evenly divided into three categories:\\
\begin{itemize}
    \item High enhancement over a short period of time; companions that cross the asteroid-belt or get near it or have a high inclination but not so high as to destabilize the system.
    \item Moderate enhancement over a moderate amount of time; companions that have a large pericentre and a low inclination. The weak Kozai-forcing combined with their large separation means that they destabilize their ``ck'' asteroids over this moderate time-scale and then the depletion slows significantly.
    \item Low enhancement over a long time: companions that have a moderate inclination and a pericentre greater than 90 or so au. These companions can deplete most if not all of the belt given sufficient time and it results in an enhancement by a factor of 2-10 on a time-scale of hundreds of Myr to Gyrs. 
\end{itemize}

We see here that the type of Companions that we concluded pose the greatest danger for life on a habitable planet when discussing the initial 1 Gyr simulations corresponding to a considerable fraction of possible Companion orbits. That is, the low inclination Companions (25 and 40 degrees but for more distant pericentres also the 50 degree) as they can produce an enhancement that persists for a much longer time.

\section{Conclusions}\label{sec:concs}
Using a simple model of a planetary system with a planet in its habitable zone, we have studied the impact that capturing a planet on a wide, inclined orbit would have on the habitability of the system by considering the delivery of impactors. From this simple model we then generalized further and considered the possible outcomes of capturing planets. Our key findings are: 
\begin{itemize}
    \item We show that due to the planetary system not being in the test particle limit, the Kozai-oscillations deviate somewhat from the analytic description of the problem~\citep{Innanen1997}, as we can see in figure~\ref{fig:kztest}. Regardless, we use the theory to determine the fraction of asteroids that cross the orbit of jupiter (figure~\ref{fig:inc_init}) and show that this is a valid approximation for what happens in the first 10 Myr of the system, for all Companion inclinations; as shown in table~\ref{tab:Time_limitied_fracs}.
    \item For the SEjC system (Sun+Earth+Jupiter-mass planet at 10 au + Jupiter-mass Companion at 100 au) we find that for all inclinations roughly 4 out of $10^6$ asteroids that are destabilized will hit the Earth (figures~\ref{fig:norm_error} and~\ref{fig:fit}), which when considering the current Kuiper belt is equivalent to $\sim10^4$ km-sized impacts.
    \item On a 10 Myr time-scale after capture we see very high impact rates for the Earth from the highly inclined captures, thousands of times higher than the present rate (figure~\ref{fig:impacts}). A Companion that was captured in a close encounter in the Sun's birth cluster and unbound by a later encounter could have produced an impact rate equivalent to the Late Heavy Bombardment in the early stages of the Solar system.
    \item For habitable planets Companions captured on low inclination orbits have worse consequences, as it can give an enhanced impact rate for several Gyr (figure~\ref{fig:deptime}), rather than the short cataclysmic event produced by highly inclined captured companion.
    \item Inclined captured companions are very effective at clearing out the asteroid belt, with Companions that have an inclination greater than 55 degrees ($\sim$60\% of an isotropic distribution) and pericentre less than 100 au eventually clearing out the whole belt. This is contingent on the precession from the inner planets on the belt being slower than the Kozai oscillations (compare fig~\ref{fig:regimes} and~\ref{fig:deptime})
    \item With $e\sim0.7$ for captured planets as it follows a thermal distribution (see e.g.~\citealp{Li2019}) the pericentres will likely often be small and captured planets will often remove a large fraction of the belts. 
    \item We find that the time-scales on which the belt is cleared out can vary from a couple of tens of Myr to  Gyrs (figures~\ref{fig:deptime} and~\ref{fig:depletion}).
\end{itemize}
\section*{Acknowledgments}
The authors are supported by the project grant 2014.0017 ``IMPACT" from the Knut and Alice Wallenberg Foundation. AJM acknowledges support from the Swedish Research Council (starting grant 2017-04945) and from the Swedish National Space Agency (career grant 120/19C). The computations were enabled by resources provided by the Swedish National Infrastructure for Computing (SNIC) at LUNARC partially funded by the Swedish Research Council through grant agreement no. 2016-07213.

\section*{Data Availability}
The data underlying this article will be shared on reasonable request
to the corresponding author

\bibliography{Prefs.bib}
\bibliographystyle{mnras}
\newpage
\appendix

\end{document}